\documentclass[reqno,12pt]{amsart}
\usepackage{graphics}
\usepackage{amssymb,amsmath}
\input cyracc.def
\font\tencyr=wncyr8
\def\cyr{\tencyr\cyracc}
\newsavebox{\ukrB}
\savebox{\ukrB}{{\cyr B}}
\newtheorem{theorem}{Theorem}[section]
\newtheorem{lemma}[theorem]{Lemma}
\newtheorem{definition}[theorem]{Definition}
\newtheorem{remark}[theorem]{Remark}
\newcounter{claim}
\renewcommand{\theclaim}{\Alph{claim}}
\newenvironment{claim}{\refstepcounter{claim}%
\par\medskip\par\noindent{\it Claim~\theclaim.~}~\rm}%
{\par\smallskip\par}
\newenvironment{subproof}{\par\noindent{\sl Proof of Claim~\theclaim.~}}%
{$\,\triangleleft$\par\medskip\par}
\newenvironment{bfenumerate}%
{%
\begin{enumerate}}{\end{enumerate}}
\newcounter{oq}
\newcommand{\que}{\refstepcounter{oq}\par{\bf \theoq.}~}
\newcommand{\fix}[1]{\mathit{fix}(#1)}
\newcommand{\reals}{\mathbb{R}}
\newcommand{\function}[2]{:#1 \rightarrow #2}
\newcommand{\of}[1]{\left( #1 \right)}
\newcommand{\setdef}[2]{\left\{ \hspace{0.5mm} #1 : \hspace{0.5mm} #2 \right\}}
\newcommand{\refeq}[1]{(\ref{eq:#1})}
\newcommand{\eps}{\varepsilon}

\newcommand{\EE}[1]{ {\mathbb E} \left[ #1 \right] }
\newcommand{\PP}[1]{ {\mathbb P} \left[ #1 \right] }
\newcommand{\fixx}[1]{\mathit{fix}^X(#1)}
\newcommand{\FIX}[1]{\mathit{FIX}(#1)}
\newcommand{\Fix}[1]{\mathit{Fix}(#1)}
\newcommand{\fixl}[1]{\mathit{fix}^-(#1)}
\newcommand{\calS}{{\mathcal S}}
\newcommand{\colX}{{\mathcal X}}
\newcommand{\colZ}{{\mathcal Z}}
\newcommand{\calH}{{\mathcal H}}
\newcommand{\cspace}{{\mathrm F}(X)}
\newcommand{\bspace}{{\mathrm B}(X)}
\newif\ifnotesw\noteswtrue

\newcommand{\edit}[1]{\ifnotesw \marginpar%
 [{\scriptsize\it\begin{minipage}[t]{\marginparwidth}\raggedleft#1\end{minipage}}]%
 {\scriptsize\it\begin{minipage}[t]{\marginparwidth}\raggedright#1\end{minipage}} \fi}

\title{Untangling planar graphs\\ from a specified vertex position
--- Hard cases}

\author{M.~Kang}
\address{Institut f\"ur Informatik, Humboldt Universit\"at zu Berlin, 
D-10099 Berlin}

\author{O.~Pikhurko$^*$}
\address{Department of Mathematical Sciences,
Carnegie Mellon University, Pittsburgh, PA 15213}
\thanks{$^*$\,Partially supported by the National Science Foundation through Grant DMS-0758057
and an Alexander von Humboldt fellowship.}

\author{A.~Ravsky}
\address{Institute for Applied Problems of Mechanics and Mathematics, 
Naukova St.\ 3{\cyr B}, Lviv 79060, Ukraine}

\author{M.~Schacht}
\address{Institut f\"ur Informatik, Humboldt Universit\"at zu Berlin, 
D-10099 Berlin}

\author{O.~Verbitsky\,$^\dag$}
\address{Institute for Applied Problems of Mechanics and Mathematics, 
Naukova St.\ 3{\cyr B}, Lviv 79060, Ukraine}
\thanks{$^\dag$\,Supported by an Alexander von Humboldt return fellowship.}

\date{22 November 2010}

\begin{document} 

\begin{abstract}
Given a planar graph $G$,
we consider drawings of $G$ in the plane where edges are represented
by straight line segments (which possibly intersect).
Such a drawing is specified by an injective embedding $\pi$
of the vertex set of $G$ into the plane.
Let $\fix{G,\pi}$ be the maximum integer $k$  such that there exists
a crossing-free redrawing $\pi'$ of $G$ which keeps $k$ vertices
fixed, i.e., there exist $k$ vertices $v_1,\dots,v_k$ of $G$ such that 
$\pi(v_i)=\pi'(v_i)$ for $i=1,\dots,k$. Given a set of points $X$, let
$\fixx G$ denote the value of $\fix{G,\pi}$ minimized over $\pi$
locating the vertices of $G$ on~$X$.
The absolute minimum of $\fix{G,\pi}$ is denoted by~$\fix G$.

For the wheel graph $W_n$, we prove that $\fixx{W_n}\le(2+o(1))\sqrt n$
for every $X$. With a somewhat worse constant factor this is as well true
for the fan graph $F_n$. 
We inspect also other graphs for which it is known that $\fix{G}=O(\sqrt n)$.

We also show that the minimum value $\fix G$ of the parameter
$\fixx G$ is always attainable by a collinear~$X$.
\end{abstract}

\maketitle

\markleft{\sc M.~KANG, O.~PIKHURKO, A.~RAVSKY, M.~SCHACHT, AND O.~VERBITSKY}
\markright{\sc UNTANGLING PLANAR GRAPHS --- HARD CASES}

\section{Introduction}\label{s:intro}
\subsection{The problem of untangling a planar graph}
In a \emph{plane graph}, each vertex $v$ 
is a point in  $\reals^{2}$ and each edge $uv$ 
is represented as a continuous plane curve with endpoints $u$ and $v$.
All such curves are supposed to be non-self-crossing and any two of them
either have no common point or share a common endvertex.
An underlying abstract graph of a plane graph is called \emph{planar}.
A planar graph can be drawn as a plane graph in many ways, and
the Wagner-F\'ary-Stein theorem (see, e.g., \cite{NRa})
states that there always exists a \emph{straight line drawing}
in which every edge is represented by a straight line segment.

Let $V(G)$ denote the vertex set of a planar graph $G$. In this paper,
by a \emph{drawing} of $G$ we mean an arbitrary injective map
$\pi\function{V(G)}{\reals^2}$. We suppose that each edge $uv$
of $G$ is drawn as the straight line segment with endpoints $\pi(u)$
and $\pi(v)$.
Due to possible edge crossings and even overlaps,
$\pi$ may not be a plane drawing of~$G$. Hence it is natural to ask:
\begin{quote}
How many vertices have to be moved to obtain from $\pi$\\
a plane (i.e., crossing-free) straight line drawing of~$G$?
\end{quote}
Alternatively, we could 
allow in $\pi$ curved edges without their exact specification; such a drawing 
could be always assumed to be a plane graph. Then our task would be
to \emph{straighten} $\pi$ rather than eliminate edge crossings.

More formally, for a planar graph $G$ and a drawing $\pi$, let
$$
\fix{G,\pi}=\max_{\pi'}|\setdef{v\in V(G)}{\pi'(v)=\pi(v)}|
$$
where the maximum is taken over all plane straight line drawings $\pi'$ of $G$.
Furthermore, let
\begin{equation}\label{eq:deffix}
\fix G=\min_\pi\fix{G,\pi}.
\end{equation}
In other words, $\fix G$ is the maximum number of vertices
which can be fixed in any drawing of $G$ while \emph{untangling} it.

No efficient algorithm determining the parameter $\fix G$ is known.
Moreover, computing $\fix{G,\pi}$ is known to be NP-hard \cite{merged,Ver}.

Improving a result of Goaoc et al.~\cite{merged},
Bose et al.~\cite{Bose} showed that
$$
\fix G\ge(n/3)^{1/4}
$$
for every planar graph $G$, where here and in the rest of this paper~$n$ 
denotes the number of vertices
in the graph under consideration. Better bounds on $\fix G$ are known
for cycles~\cite{PTa}, trees~\cite{merged,Bose} and, more generally, outerplanar 
graphs~\cite{merged,RVe}. In all these cases it was shown that $\fix G=\Omega(n^{1/2})$.
For cycles Cibulka \cite{Cib} proves a better lower bound of $\Omega(n^{2/3})$.

Here we are interested in upper bounds on $\fix G$, that is, in examples of graphs
with small $\fix G$. Moreover, let $X$ be an arbitrary set of $n$ points in the 
plane and define
$$
\fixx G=\min_\pi\setdef{\fix{G,\pi}}{\pi(V(G))=X}.
$$
Note that $\fix G=\min_X\fixx G$.
This notation allows us to formalize another natural question.
Can untangling of a graph become easier if the set $X$ of vertex positions
has some special properties (say, if it is known that $X$ is \emph{collinear}, i.e.,
lies on a line, or is in \emph{convex position}, i.e., 
no $x\in X$ lies in the convex hull of $X\setminus\{x\}$)? 
This question admits several variations:
\begin{itemize}
\item
For which $X$ can one attain equality $\fixx G=\fix G$?
\item
Are there graphs with $\fixx G$ small for \emph{all}~$X$?
\item
Are there graphs such that $\fixx G$ is for some $X$ considerably
larger than $\fix G$?
\end{itemize}

\subsection{Prior results}\label{ss:prior}
The \emph{cycle} (resp.\ \emph{path}; \emph{empty graph})
on $n$ vertices will be denoted by $C_n$ (resp.\ $P_n$; $E_n$).
Recall that the \emph{join} of vertex-disjoint graphs $G$ and $H$
is the graph $G*H$ consisting of the union of $G$ and $H$ and all
edges between $V(G)$ and $V(H)$. The graphs $W_n=C_{n-1}*E_1$ (resp.\
$F_n=P_{n-1}*E_1$; $S_n=E_{n-1}*E_1$) are known as \emph{wheels} (resp.\
\emph{fans}; \emph{stars}). By $kG$ we denote
the disjoint union of $k$ copies of a graph~$G$.

Pach and Tardos~\cite{PTa} were first who established a principal fact:
Some graphs can be drawn so that, in order to untangle them, one has
to shift almost all their vertices. In fact, 
this is already true for cycles.
More precisely,  Pach and Tardos~\cite{PTa} proved that
\begin{equation}\label{eq:PTa}
\fixx{C_n}=O((n\log n)^{2/3})\mathrm{\ for\ any\ }X\mathrm{\ in\ convex\ position}.
\end{equation}

The best known upper bounds are of the form $\fix G=O(\sqrt n)$.
Goaoc et al.\ \cite{Goaoc}\footnote{%
The conference presentations \cite{Goaoc} and \cite{SWo} were subsequently combined
into the journal paper~\cite{merged}.}
showed it for certain triangulations. 
More specifically, they proved that
\begin{equation}\label{eq:Goaoc}
\fixx{P_{n-2}*P_2}<\sqrt n+2\mathrm{\ for \ any\ collinear\ }X.
\end{equation}

Shortly after \cite{Goaoc} and independently of it,
there appeared our manuscript \cite{KSchV}, which was actually 
a starting point of the current paper. For infinitely many $n$,
we constructed a family $\calH_n$ of 3-connected
planar graphs on $n$ vertices with $\max_{H\in\calH_n}\fix{H}=o(n)$. Though no explicit bound was
specified in \cite{KSchV}, a simple analysis of our construction
reveals that
\begin{equation}\label{eq:fix}
\fixx{H_n}\le2\sqrt n+1\mathrm{\ for\ any\ }X\mathrm{\ in\ convex\ position,}
\end{equation}
where $H_n$ denotes an arbitrary member of $\calH_n$.
While the graphs in $\calH_n$ are not as simple as $P_{n-2}*P_2$ and the subsequent examples in the
literature, the construction of $\calH_n$ has the advantage that this class contains graphs with
certain special properties, such as bounded vertex degrees.
By a later result of Cibulka \cite{Cib}, we have $\fix G=O(\sqrt n(\log n)^{3/2})$
for every $G$ with maximum degree and diameter bounded by a logarithmic function. 
Note in this respect that $\calH_n$ contains graphs with bounded maximum degree that
have diameter~$\Omega(\sqrt n)$.

In subsequent papers \cite{SWo,Bose} examples of graphs with small $\fix G$
were found in special classes of planar graphs, such
as outerplanar and even acyclic graphs.
Spillner and Wolff \cite{SWo} showed for the fan graph that
\begin{equation}\label{eq:fans}
\fixx{F_n}<2\sqrt n+1\mathrm{\ for \ any\ collinear\ }X
\end{equation}
and Bose et al.\ \cite{Bose} established for the star forest with $n=k^2$ vertices that
\begin{equation}\label{eq:stars}
\fixx{kS_k}\le3\sqrt n-3\mathrm{\ for \ any\ collinear\ }X.
\end{equation}
Finally, Cibulka \cite{Cib} proved that
$$
\fixx{G}=O((n\log n)^{2/3})\mathrm{\ for\ any\ }X\mathrm{\ in\ convex\ position}
$$
for all 3-connected planar graphs.

\subsection{Our present contribution}\label{ss:contrib}
In Section \ref{s:collin} we notice that the choice of a collinear
vertex position in \refeq{Goaoc}, \refeq{fans}, and \refeq{stars}
is actually optimal for proving upper bounds on $\fix G$.
Specifically, we show that for any $G$ the equality $\fixx G=\fix G$
is attained by some collinear $X$ (see Theorem \ref{thm:fixlfix}).

In Section \ref{s:fixx} we extend the bound
$\fix G=O(\sqrt n)$ in the strongest way with respect to specification
of vertex positions. We prove that
\begin{eqnarray}
\fixx{W_n}&\le&\ \,(2+o(1))\sqrt n\mathrm{\ \ for\ every\ }X, \label{eq:fixx}\\
\fixx{F_n}&\le&(2\sqrt2+o(1))\sqrt n\mathrm{\ \ for\ every\ }X \label{eq:fixx2}
\end{eqnarray}
(see Theorem \ref{thm:FIXWnFn}).
Let us define
$$
\FIX G=\max_X\fixx G
$$
(while $\fix G=\min_X\fixx G$).
With this notation, \refeq{fixx} and \refeq{fixx2} read
$$
\FIX{W_n}\le(2+o(1))\sqrt n\quad\mbox{and}\quad\FIX{F_n}\le(2\sqrt2+o(1))\sqrt n.
$$

\begin{figure}
\centerline{\includegraphics{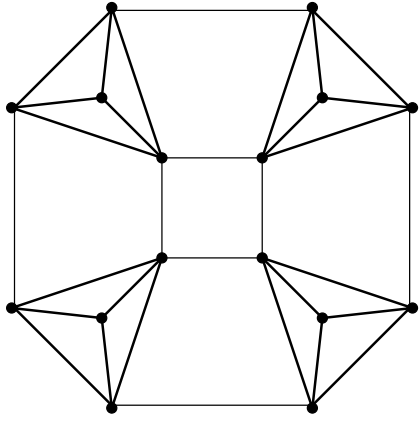}}
\caption{Example of a graph in $\calH_{16}$.}
\label{fig:Gk4}
\end{figure}

In Section \ref{s:hulls} we discuss an approach attempting to give
an analog of \refeq{fixx} for the aforementioned family of graphs
$\calH_n$. A member of this family is defined as a plane graph of the
following kind. 
Let $k\ge3$ and $n=k^2$. Draw $k$ triangulations, each having $k$ vertices,
so that none of them lies inside an inner face of any other
triangulation. Connect these triangulations by some more edges making
the whole graph 3-connected.
$\calH_n$ is the set of all 3-connected planar graphs obtainable in this way.
This set is not empty. Indeed, we can allocate the $k$ triangulations in a
cyclic order and connect each neighboring pair by two vertex-disjoint edges
as shown in Fig.~\ref{fig:Gk4}. Note that $k$ new edges form a cycle $C_k$
and the other $k$ new edges participate in a cycle $C_{2k}$. 
If we remove any two vertices from the obtained graph,
each triangulation as well as the whole ``cycle'' stay connected
(since the aforementioned cycles $C_k$ and $C_{2k}$ are vertex-disjoint,
at most one of them can get disconnected).

Note that, if we start with triangulations with bounded vertex degrees, 
the above construction gives us a graph with bounded maximum degree.
In this situation our argument for \refeq{fixx}
does not work. We hence undertake a different approach.

Given a set of colored points in the plane, we
call it \emph{clustered} if its monochromatic parts have pairwise disjoint
convex hulls. Given a set $X$ of $n=k^2$ points, let $C(X)$ denote
the maximum cardinality of a clustered subset existing in $X$
under any balanced coloring of $X$ in $k$ colors (see Definition \ref{def:CX}).
It is not hard to show (see Lemma \ref{lem:fixxC}) that
\begin{equation}\label{eq:HnC}
\fixx{H_n}\le C(X)+k,
\end{equation}
where $H_n$ denotes an arbitrary graph in $\calH_n$.
We prove that $C(X)=O(n/\log n)$ for every $X$, which implies that $\FIX{H_n}=O(n/\log n)$ 
(Theorem~\ref{thm:FIXHn}).

Better upper bounds for $C(X)$ would give us better upper bounds for $\FIX{H_n}$.
Note that $C(X)$ has relevance also to the star forest ${kS_k}$, namely
\begin{equation}\label{eq:kSkC}
\fixx{kS_k}\ge C(X)-k
\end{equation}
(see part 2 of Lemma \ref{lem:fixxC}).
Thus, if there were a set $X$ with $C(X)\gg k$, the parameter $\FIX{kS_k}$
would be far apart from $\fix{kS_k}$.

As we do not know how close or far away the parameters $\fix G$ and $\FIX G$ are
for $G=H_n$ and $G=kS_k$, the two graph families deserve further attention.
Section \ref{s:weak} is devoted to estimation of $\fixx G$ for $X$ in
\emph{weakly convex position}, which means that the points in $X$
lie on the boundary of a convex body (including the cases that $X$
is in convex position and that $X$ is a collinear set). Since
$C(X)<2k$ for any $X$ in weakly convex position, by \refeq{HnC} we
obtain $\fixx{H_n}<3\sqrt n$ for such $X$ (Theorem~\ref{thm:fix}). 

This result for $\calH_n$ together with the stronger results obtained for
$W_n$ and $F_n$ in Section \ref{s:fixx} might suggest
that $\fixx G=O(\fix G)$ should hold for any $G$ whenever $X$ is in weakly convex position.
The simplest case where we are not able to confirm this conjecture is $G=kS_k$.
By \refeq{HnC} and \refeq{kSkC} we have $\fixx{H_n}\le\fixx{kS_k}+2k$ for any $k$ and $n=k^2$,
and bounding $\fixx{kS_k}$ from above seems harder.
Nevertheless, even here we have a rather tight bound: 
If $X$ is in weakly convex position, then
$\fixx{kS_k}=O(\sqrt n\,2^{\alpha(\sqrt{n})})$, where $\alpha(\cdot)$
denotes the inverse Ackermann function (Theorem~\ref{thm:stars}).

We conclude with a list of open questions in Section~\ref{s:open}.

\section{Hardness of untangling from a collinear position}\label{s:collin}

\begin{theorem}\label{thm:fixlfix}
For every planar graph $G$ we have $\fix G=\fixx G$ for some collinear~$X$.
\end{theorem}
Theorem~\ref{thm:fixlfix} can
be deduced from \cite[Lemma 1]{Bose}. For the reader's convenience,
we give a self-contained proof.

\begin{proof} 
Let $\fixl G$ denote the minimum value of $\fixx G$ over collinear $X$.
We have $\fix G\le\fixl G$ by definition.
The theorem actually states the converse inequality $\fix G\ge\fixl G$.
That is, given an arbitrary drawing $\pi\function{V(G)}{\reals^2}$, we have to
show that it can be untangled while keeping at least $\fixl G$ vertices fixed.

Choose Cartesian coordinates in the plane so that $\pi(V(G))$ is located between
the lines $y=0$ and $y=1$. Let $p_x,p_y:\reals^2\to\reals$ denote the projections 
onto the $x$-axis and the $y$-axis, respectively. We also suppose that the axes
are chosen so that the map $\lambda=p_x\pi$ is injective. 
Let us view $\lambda$ as a drawing of $G$, aligning all the vertices on the line $y=0$.
By definition, there is a plane drawing $\lambda'$ of $G$ such that 
the set of fixed vertices $F=\setdef{v\in V(G)}{\lambda'(v)=\lambda(v)}$
has cardinality at least $\fixl G$. 

Given a set $A\subset\reals^2$ and a real $\eps>0$, let $N_\eps(A)$
denote the $\eps$-neighborhood of $A$ in the Euclidean metric. 
For each pair of disjoint edges $e, e'$
in $\lambda'$, there is an $\eps$ such that $N_\eps(e)\cap N_\eps(e')=\emptyset$.
Since $G$ is finite, we can assume that the latter is true with the same
$\eps$ for all disjoint pairs $e, e'$. 

We now define a drawing $\pi'\function{V(G)}{\reals^2}$
by setting
$$\pi'(v)=
\begin{cases}
(p_x\pi(v),\eps p_y\pi(v)) & \mathrm{if}\ v\in F,\\
\lambda'(v) & \mathrm{otherwise}.
\end{cases}
$$
Note that $\pi'(v)\in N_\eps(\lambda'(v))$ for every $v\in V(G)$.
Since $\lambda'$ is crossing-free, so is~$\pi'$.

Finally, define a linear transformation of the plane by $a(x,y)=(x,\eps^{-1}y)$
and consider $\pi''=a\pi'$. Clearly, $\pi''$ is a plane drawing of $G$ and all
vertices in $F$ stay fixed under the transition from $\pi$ to~$\pi''$.
\end{proof}

\section{Hardness of untangling from every vertex position}\label{s:fixx}

In Section \ref{ss:monoton} we state known results on the longest monotone subsequences
in a random permutation. These results are used in Section \ref{ss:FIX} for
proving upper bounds on $\FIX{W_n}$ and $\FIX{F_n}$.

\subsection{Monotone subsequences in a random permutation}\label{ss:monoton}

By a \emph{permutation} of $[N]=\{1,2,\ldots,N\}$ we will mean a sequence
$S=s_1s_2\ldots s_N$
where each positive integer $i\le N$ occurs once (that is, $S$ determines a
one-to-one map $S\function{[N]}{[N]}$ by $S(i)=s_i$).
A subsequence $s_{i_1}s_{i_2}\ldots s_{i_k}$, where $i_1<i_2<\ldots<i_k$,
is \emph{increasing} if $s_{i_1}<s_{i_2}<\cdots<s_{i_k}$.
The length of a longest increasing subsequence of $S$ will be
denoted by $\ell(S)$. 

\begin{lemma}\label{lem:random}
Let $S_N$ be a uniformly random permutation of $\{1,2,\ldots,N\}$.

\noindent
\begin{bfenumerate}
\item \textbf{(Pilpel \cite{Pil})}
$
\EE{\ell(S_N)}\le\sum_{i=1}^N 1/\sqrt i\le2\sqrt N-1
$.

\smallskip

\item \textbf{(Frieze \cite{Fri}, Bollob\'as-Brightwell \cite{BBr})}
For any real $\epsilon>0$ there is a $\beta=\beta(\epsilon)>0$ such that
for all $N\ge N(\epsilon)$ we have
$$
\PP{\ell(S_N)\ge\EE{\ell(S_N)}+N^{1/4+\epsilon}}\le\exp\of{-N^\beta}.
$$
\end{bfenumerate}
\end{lemma}

\noindent
Further concentration results for $\ell(S_N)$ are obtained in \cite{Tal,BDJ}.

Lemma \ref{lem:random} shows that $\ell(S_N)\le2N^{1/2}(1+N^{-1/4+\epsilon})$
with probability at least $1-\exp\of{-N^\beta}$.
We will also need a bound for another parameter of $S_N$, roughly speaking, for
the maximum total length of two non-interweaving monotone subsequences of $S_N$.
Let us define this parameter more precisely.
A subsequence of a permutation $S$ will be called \emph{monotone} if it can 
be made increasing by shifting and/or reversing (as, for example, 21543). 
This notion is rather natural if we regard $S$ as a \emph{circular permutation}, 
i.e., $S$ is considered up to shifts.
Call two subsequences $S'$ and $S''$ of $S$ 
\emph{non-interweaving} if they have no common element and $S$ has no 
subsequence $s_{i_1}s_{i_2}s_{i_3}s_{i_4}$ with $s_{i_1},s_{i_3}$ occurring in $S'$ 
and $s_{i_2},s_{i_4}$ in $S''$. Define $\ell_2(S)$ to be the sum of 
the lengths of $S'$ and $S''$ maximized over non-interweaving monotone 
subsequences of $S$.

\begin{lemma}\label{lem:random2}
Let $S_N$ be a uniformly random permutation of $\{1,2,\ldots,N\}$. 
For any real $\epsilon>0$ there is a $\gamma=\gamma(\epsilon)>0$ such that
for all $N\ge N(\epsilon)$ we have
\begin{equation}\label{eq:ell2}
\PP{\ell_2(S_N)\ge2\sqrt2N^{1/2}+2N^{1/4+\epsilon}}\le\exp\of{-N^\gamma}.
\end{equation}
\end{lemma}

\begin{proof}
Given a sequence $S_N=s_1s_2\ldots s_N$ and a pair of indices $1\le i<j\le N$,
consider the splitting of the circular version of $S_N$ into two parts
$P_1=s_i\ldots s_{j-1}$ and $P_2=s_j\ldots s_Ns_1\ldots s_{i-1}$. Let $P'_1=s_{j-1}\ldots s_i$
and $P'_2=s_{i-1}\ldots s_1s_N\ldots s_j$ be the reverses of $P_1$ and $P_2$.
Denote 
$$
\lambda_{ij}=\max\{\ell(P_1),\ell(P'_1)\}+\max\{\ell(P_2),\ell(P'_2)\}.
$$
Note that $\ell_2(S_N)=\lambda_{ij}$ for some pair $i,j$.
Since there are only polynomially many such pairs,
it suffices to show for each $i,j$ that the inequality
\begin{equation}\label{eq:lij}
\lambda_{ij}\ge2\sqrt2N^{1/2}+2N^{1/4+\epsilon}
\end{equation}
holds with an exponentially small probability.
Denote the length of $P_k$ by $N_k$, so that $N_1+N_2=N$. For each $k=1,2$,
note that both $\ell(P_k)$ and $\ell(P'_k)$ are distributed identically to
$\ell(S_{N_k})$.

Suppose first that $N_1$ or $N_2$ is relatively small, say, $N_1\le2(\sqrt2-1)\sqrt N$.
Then \refeq{lij} implies that 
$$
\ell(P_2)\ge2N_2^{1/2}+2N_2^{1/4+\epsilon}
$$
or this estimate is true for $P'_2$. Provided $N$, and hence $N_2$, is large enough,
we conclude by Lemma \ref{lem:random} that \refeq{lij} happens with probability
at most $2\exp(-N_2^\beta)\le2\exp(-\frac12N^\beta)$.

Suppose now that $N_k>2(\sqrt2-1)\sqrt N$ for both $k=1,2$ and that $N$ is large enough.
Since
$
N_1^{1/2}+N_2^{1/2}\le2\of{\frac{N_1+N_2}{2}}^{1/2}=\sqrt2N^{1/2}
$,
the inequality \refeq{lij} entails that for $k=1$ or $k=2$ we must have
$$
\ell(P_k)>2N_k^{1/2}+N_k^{1/4+\epsilon}
$$
or this estimate must be true for $P'_k$. By Lemma \ref{lem:random}, the event \refeq{lij}
happens with probability no more than $4\exp\of{-c^\beta N^{\beta/2}}$, 
where $c=2(\sqrt2-1)$.

We see that, whatever $N_1$ and $N_2$ are, \refeq{ell2} holds for
any positive $\gamma<\beta/2$ and large enough~$N$.
\end{proof}

\subsection{Graphs with small $\FIX G$}\label{ss:FIX}

Recall that $\FIX G=\max_X\fixx G$.
If $\FIX G$ is small, this means that no special properties of the set
of vertex locations can make the untangling problem for $G$ easy.

\begin{lemma}\label{lem:fixx}
For any 3-connected planar graph $G$ on $n$ vertices with maximum
vertex degree $N=n-o(\sqrt n)$ we have
$$
\FIX G\le(2+o(1))\sqrt n.
$$
\end{lemma}

\begin{proof}
We have to prove that $\fixx G\le(2+o(1))\sqrt n$ for any set $X$ of $n$ points.
Let $X=\{x_1,\ldots,x_n\}$ and denote $X_N=\{x_1,\ldots,x_N\}$.
Given a point $p$ in the plane, we define a permutation 
$S_p$ describing the order in which the points in $X_N$ are visible from the 
standpoint $p$. 
If $p=x_s$ with $s\le N$, we take $p$ as the first visible point, that is,
let $s$ be the first index in the sequence $S_p$. Now, we look around starting
from the north in a clockwise direction and put $i$ before $j$ in $S_p$
if we see $x_i$ earlier than $x_j$. The ``north'' direction on the plane can be fixed 
arbitrarily. If $x_i$ and $x_j$ lie in the same direction from $p$,
we see the nearer point first, that is, $i$ precedes $i$ in $S_p$
whenever $x_i\in[p,x_j]$.

Define an equivalence relations $\equiv$ so that $S\equiv S'$ if $S$ and $S'$
are obtainable from one another by a shift.
Let us show that the quotient set $Q=\setdef{S_p}{p\in\reals^2}/_\equiv$ is finite and estimate its cardinality.
Suppose first that not all points in $X_N$ are collinear.
Let $L$ be the set of lines passing through at least
two points in $X_N$. After removal of all lines in $L$, the plane
is split into connected components that will be called \emph{$L$-faces}.
Any intersection point of two lines will be called an \emph{$L$-vertex}.
The $L$-vertices lying on a line in $L$ split this line into \emph{$L$-edges}.
Exactly two $L$-edges for each line are unbounded.
It is easy to see that $S_p\equiv S_{p'}$ whenever $p$ and $p'$ belong to the same
$L$-face or the same $L$-edge. It follows that $|Q|$
does not exceed the total amount of $L$-faces, $L$-edges, and $L$-vertices.

Let us express this bound in terms of $l=|L|\le{N\choose2}$. If we erase
all the unbounded $L$-edges, we obtain a crossing-free straight line drawing of
a planar graph with at most ${l\choose2}$ vertices. It has less than
$\frac32l^2-\frac32l$ edges and $l^2-l$ faces. Restoring the unbounded
$L$-edges, we see that the total number of $L$-edges is less than $\frac32l^2+\frac12l$  
and the number of $L$-faces is less than $l^2+l$. Therefore,
$$
|Q|<(l^2+l)+\of{\frac32\,l^2+\frac12\,l}+\of{\frac12\,l^2-\frac12\,l}<\frac34\,N^4.
$$
In the much simpler case of a collinear $X_N$, we have $|Q|\le N$.

Let $c$ be a vertex of $G$ with maximum vertex degree. By the Whitney theorem
on embeddability of 3-connected graphs, the neighbors of $c$ appear around $c$
in the same circular order $v_1,\ldots,v_N$ in any plane drawing of $G$.
Pick up a random permutation $\sigma$ of $\{1,\ldots,N\}$ and consider
a drawing $\pi\function{V(G)}X$ such that $\pi(v_i)=x_{\sigma(i)}$.
Let $\pi'$ be an untanglement of $\pi$. Let $p=\pi'(c)$ and denote
the set of all shifts and reverses
of the permutation $S_p$ by~$\mathcal S_p$.

We have to estimate the number of vertices remaining fixed 
under the transition from $\pi$ to $\pi'$, that is, the cardinality
of the set $F=\setdef{\pi(v)}{v\in V(G), \pi(v)=\pi'(v)}$.
Let $F^*=\setdef{\pi(v_i)\in F}{i\le N}$, which is the subset of $F$
corresponding to the fixed neighbors of $c$. Note that
$|F\setminus F^*|\le n-N$ and recall that $n-N=o(\sqrt n)$ by our assumption.
It follows that $|F|\le|F^*|+o(\sqrt n)$, and we have to estimate~$|F^*|$.

The points in $F^*$ go around $p$ in the canonical Whitney order.
This means that the indices of the corresponding vertices
form an increasing subsequence in $\sigma^{-1}S$ for some $S\in\mathcal S_p$.
For each $S$, the composition $\sigma^{-1}S$ is a random permutation of $\{1,\ldots,N\}$.
Recall that, irrespectively of the choice of $p=\pi'(c)$,
there are at most $2N|Q|<\frac32N^5$ possibilities for $S$.
By Lemma \ref{lem:random}, every increasing subsequence of $\sigma^{-1}S$
has length at most $2N^{1/2}+N^{1/4+\epsilon}$ with probability at least $1-O(N^5\exp\of{-N^\beta})$.
Thus, if $N$ is sufficiently large, we have $|F^*|\le(2+o(1))\sqrt n$ for all
untanglements $\pi'$ of some drawing $\pi$ (in fact, this is true
for almost all~$\pi$). This implies the required bound $|F|\le(2+o(1))\sqrt n$.
\end{proof}

While Lemma \ref{lem:fixx} immediately gives us a bound on $\FIX{W_n}$
for the wheel graph, this lemma does not apply directly to the fan graph $F_n$ 
because it is not 3-connected and has a number of essentially different plane drawings. 
Nevertheless, all these drawings are still rather structured, which makes
analysis of the fan graph only a bit more complicated.
Indeed, denote the central vertex of $F_n$ by $c$ and let $v_1\ldots v_{n-1}$ 
be the path of the other vertices.
Let $\alpha$ be a plane drawing of $F_n$. Label each edge $\alpha(c)\alpha(v_i)$
with number $i$ and denote the circular sequence in which the labels follow each other
around $\alpha(c)$ by $R_\alpha$. Split $R_\alpha$ into two pieces.
Let $R'_\alpha$ be the sequence of labels starting with $1$, ending with $n-1$,
and containing all intermediate labels if we go around $\alpha(c)$ clockwise.
Let $R''_\alpha$ be the counter-clockwise analog of $R'_\alpha$.
Note that $R'_\alpha$ and $R''_\alpha$ overlap in $\{1,n-1\}$.

\begin{lemma}\label{lem:fans}
Both $R'_\alpha$ and $R''_\alpha$ are monotone.
\end{lemma}

\begin{proof}
We proceed by induction on $n$. The base case of $n=3$ is obvious.
Suppose that the claim is true for all plane drawings of $F_n$
and consider an arbitrary plane drawing $\alpha$ of $F_{n+1}$.
Let $\beta$ be obtained from $\alpha$ by erasing $\alpha(v_n)$
along with the incident edges. Obviously, $\beta$ is a plane drawing of $F_{n}$.

In the drawing $\alpha$ of $F_{n+1}$, we consider the triangle
$T$ with vertices $\alpha(c)$, $\alpha(v_{n-1})$, and
$\alpha(v_n)$. Clearly, all points $\alpha(v_i)$ for $i\le n-2$ are inside $T$
or all of them are outside. In both cases, $n-1$ and $n$ are neighbors in $R_\alpha$.
Therefore, $R_\alpha$ is obtainable from $R_\beta$ by inserting $n$ on the one
or the other side next to $n-1$. It follows that $R'_\alpha$ is obtained from
$R'_\beta$ either by appending $n$ after $n-1$ or by replacing $n-1$ with $n$
(the same concerns $R''_\alpha$ and $R''_\beta$). It remains to note that
both operations preserve monotonicity.
\end{proof}

We are now prepared to obtain upper bounds on $\FIX G$ for the wheel graph $W_n$ 
and the fan graph $F_n$.
Note that, up to a small constant factor, these bounds
match the lower bound $\fix{F_n}\ge\fix{W_n}\ge\sqrt{n-2}$ (which follows, e.g., from
\cite[Theorem 4.1]{RVe}).

\begin{theorem}\label{thm:FIXWnFn}
\mbox{}

\begin{bfenumerate}
\item
$\FIX{W_n}\le(2+o(1))\sqrt n$.
\item
$\FIX{F_n}\le(2\sqrt2+o(1))\sqrt n$.
\end{bfenumerate}
\end{theorem}
 
\begin{proof}
The bound for $W_n$ follows directly from Lemma \ref{lem:fixx} as observed before.

As for $F_n$, notice that the argument of Lemma \ref{lem:fixx} becomes applicable
if, in place of the Whitney theorem, we use Lemma \ref{lem:fans}.
Let $\pi$ be a random location of $V(F_n)$ on $X$, as in the proof
of Lemma \ref{lem:fixx}. 
More precisely, let $v_1\ldots v_{n-1}$ denote the path of non-central vertices
in $F_n$. We pick a random permutation $\sigma$ of $\{1,\ldots,n-1\}$ and set
$\pi(v_i)=x_{\sigma(i)}$. As established in the proof of Lemma \ref{lem:fixx},
the set $X$ determines a set of permutations $\calS_X$ with $|\calS_X|=O(n^4)$
such that, from any standpoint $p$ in the plane, the vertices 
$v_1,\ldots, v_{n-1}$ are visible in the circular order $\tau_p=\sigma^{-1}S$
for some $S\in\calS_X$.

Let $\alpha$ be any untangling of $\pi$ and $R_\alpha$
be the associated order on the neighborhood of the central vertex $\alpha(c)$.
By Lemma \ref{lem:fans}, $R_\alpha$ consists of two monotone parts
$R'_\alpha$ and $R''_\alpha$. The set $F$ of fixed vertices is correspondingly
split into $F'$ and $F''$.
Since $R'_\alpha$ and $R''_\alpha$ overlap in two elements,
$F'$ and $F''$ can have one or two common vertices. If this happens,
we remove those from $F''$.
Notice that the indices of the vertices in $F'$ and in $F''$ form
non-interweaving monotone subsequences of $\tau_{\alpha(c)}$. 
Therefore, $|F'|+|F''|\le\ell_2(\tau_{\alpha(c)})$
and part 2 of the theorem follows from Lemma~\ref{lem:random2}.
\end{proof}

\section{Making convex hulls disjoint}\label{s:hulls}

In Section \ref{ss:prior} we listed the few graphs for which an upper
bound $\fix G=O(\sqrt n)$ is known, namely $P_{n-2}*P_2$, $F_n$, $H_n\in\calH_n$, and $kS_k$.
By Theorem \ref{thm:FIXWnFn} in the former two cases we have a stronger
result $\FIX G=O(\sqrt n)$ (note that $P_{n-2}*P_2$ contains $W_n$ as a subgraph).
We now consider a problem related to estimating the parameters $\FIX{H_n}$
and $\FIX{kS_k}$. 

\begin{definition}\label{def:CX}\rm
Let $n=k^2$ and $X$ be an $n$-point set in the plane.
Given a partition $X=X_1\cup\ldots\cup X_k$, we regard $\colX=\{X_1,\ldots,X_k\}$ 
as a coloring of $X$ in $k$ colors.
We will consider only \emph{balanced} $\colX$ with each $|X_i|=k$.
Call a set $Y\subseteq X$ \emph{clustered} if the monochromatic classes
$Y_i=Y\cap X_i$ have pairwise disjoint convex hulls.
Let $C(X,\colX)$ denote the largest size of a clustered subset of $X$.
Finally, define $C(X)=\min_\colX C(X,\colX)$.
\end{definition}

\begin{lemma}\label{lem:fixxC}
Let $H_n$ denote an arbitrary graph in $\calH_n$.
\begin{bfenumerate}
\item
$\fixx{H_n}\le C(X)+k$.
\item
$\fixx{kS_k}\ge C(X)-k$.
\end{bfenumerate}
\end{lemma}

\begin{proof}
{\sl 1.}
Recall that $H_n$ is defined as a plane graph whose vertex set
$V(H_n)=V_1\cup\ldots\cup V_k$ is partitioned so that each $V_i$ spans
a triangulation and these $k$ triangulations are in the outer faces of each other.
Take $\colX$ such that $C(X,\colX)=C(X)$ and
$\pi\function{V(H_n)}{X}$ such that $\pi(V_i)=X_i$.
Consider an untanglement $\pi'$ of $\pi$ and denote the set
of fixed vertex locations by $Y$. 
By the Whitney theorem, $\pi'$ is obtainable from the plane graph $H_n$
by a homeomorphism of the plane, possibly after turning some inner face of $H_n$
into the outer face.
Since $V_i$ spans a triangulation in $H_n$, the convex hull of $\pi'(V_i)$
is a triangle $T_i$. Since the corresponding triangulations are pairwise
disjoint in $H_n$, the triangles $T_i$'s are pairwise disjoint possibly with
a single exception for some $T_s$ containing all the other triangles.
Let $Y_i=Y\cap X_i$. It follows that the convex hulls of the
$Y_i$'s do not intersect, perhaps with an exception for a single
set $Y_s$. The exception may occur if $\pi'$ is homeomorphic to a version 
of $H_n$ with different outer face.
Therefore, $|Y|\le C(X)+k$, where the term $k$ corresponds to the exceptional~$Y_s$.

{\sl 2.}
Given an arbitrary drawing $\pi\function{V(kS_k)}{X}$ of the star forest, 
we have to untangle it while keeping at least $C(X)-k$ vertices
fixed. Let $V(kS_k)=V_1\cup\ldots\cup V_k$ where each $V_i$ is the vertex set
of a star component. Define a coloring $\colX$ of $X$ by $X_i=\pi(V_i)$.
Let $Y$ be a largest clustered subset of $X$. 
Choose pairwise disjoint open convex sets $C_1,\ldots,C_k$ so that $C_i$ 
contains $Y_i=Y\cap X_i$ for all $i$. Redraw $kS_k$ so that, for each $i$, 
the $i$-th star component is contained in $C_i$. It is
clear that, doing so, we can leave all non-central vertices in $Y$ fixed.
Thus, we have at least $|Y|-k\ge C(X)-k$ fixed vertices.
\end{proof}

\begin{lemma}\label{lem:CX}
For any set $X$ of $n=k^2$ points in the plane, we have $C(X)=O(n/\log n)$.
\end{lemma}

\begin{proof}
Let $\bspace$ denote the set of all balanced $k$-colorings of $X$, i.e.,
the set of partitions $X=X_1\cup\ldots\cup X_k$ with each $|X_i|=k$.
We have $|\bspace|=n!/(k!)^k$.

Call a $k$-tuple of subsets $Z_1,\ldots,Z_k\subset X$ a \emph{crossing-free
coloring} of $X$ if the $Z_i$'s have pairwise disjoint convex hulls.
We do not exclude that some $Z_i$'s are empty and the coloring is partial, i.e.,
$\bigcup_{i=1}^kZ_i\subsetneq X$. Denote the set of all crossing-free colorings of $X$
by~$\cspace$.

Let $\colX\in\bspace$. An estimate $C(X,\colX)\ge a$ means that
\begin{equation}\label{eq:XZ}
\sum_{i=1}^k|X_i\cap Z_i|\ge a
\end{equation}
for some $\colZ\in\cspace$. Regard $\colX$ and $\colZ$ as elements of the space
$\{1,\ldots,k,k+1\}^X$ of $(k+1)$-colorings of $X$, where the new color $k+1$ is
assigned to the points that are uncolored in $\colZ$. Then \refeq{XZ} means
that the Hamming distance between $\colX$ and $\colZ$ does not exceed $n-a$.
Note that the $(n-a)$-neighborhood of $\colZ$ can contain no more than 
${n\choose n-a}k^{n-a}$ elements of $\bspace$. Therefore, an estimate $C(X)<a$
would follow from inequality
\begin{equation}\label{eq:CB}
|\cspace|{n\choose a}k^{n-a} < |\bspace|.
\end{equation}

Given a partition $Z=P_1\cup\ldots\cup P_m$ of a point set $Z$,
we call it \emph{crossing-free} if the
convex hulls of the $P_i$'s are nonempty and pairwise disjoint. 
According to Sharir and Welzl \cite[Theorem 5.2]{ShW},
the overall number of crossing-free partitions of any $l$-point set
$Z$ is at most $O(12.24^l)$.
In order to derive from here a bound for the number of
crossing-free \emph{colorings}, with each coloring $(Z_1,\ldots,Z_k)$
we associate a partition $(P_1,\ldots,P_m)$ of the union $Z=\bigcup_{i=1}^kZ_i$
so that $(P_1,\ldots,P_m)$
is the result of removing all empty sets from the sequence $(Z_1,\ldots,Z_k)$.
Since $(P_1,\ldots,P_m)$ is the crossing-free partition of a subset of $X$,
the Sharir-Welzl bound implies that the number of all possible partitions $(P_1,\ldots,P_m)$
obtainable in this way does not exceed $O(24.48^n)$.
Since $(Z_1,\ldots,Z_k)$ can be restored from $(P_1,\ldots,P_m)$ in
${k\choose m}$ ways, we obtain
$|\cspace| < c\, 2^k 24.48^n$
for a constant $c$. Thus, we would have \refeq{CB} provided
$$
c\, 2^k 24.48^n\frac{n^a}{a!}k^{n-a} \le \frac{n!}{(k!)^k}.
$$
Taking logarithm of both sides, we see that the latter inequality
holds for all sufficiently large $n$ if we set $a=6.4\,n/\ln n$.
\end{proof}

Part 1 of Lemma \ref{lem:fixxC} and Lemma \ref{lem:CX} immediately give us
the main result of this section.

\begin{theorem}\label{thm:FIXHn}
$\FIX{H_n}=O(n/\log n)$ for an arbitrary $H_n\in\calH_n$.
\end{theorem}

\noindent
Note that the bound of Theorem \ref{thm:FIXHn} is the best upper bound on
$\FIX G$ that we know for graphs with bounded vertex degrees.

\section{Hardness of untangling from weakly convex position}\label{s:weak}

Despite the observations made in Section \ref{s:hulls}, we do not know
whether or not $\fixx{H_n}$ and $\fixx{kS_k}$ are close to, respectively, 
$\fix{H_n}$ and $\fix{kS_k}$ for every location $X$ of the vertex set.
We now restrict our attention to point sets $X$ in weakly convex position,
i.e., on the boundary of a convex plane body.

We will use Davenport-Schinzel sequences defined as follows 
(see, e.g.,~\cite{ASh} for more details).
An integer sequence $S=s_1\ldots s_n$ is called a \emph{$(k,p)$-Davenport-Schinzel
sequence} if the following conditions are met:
\begin{itemize}
\item
$1\le s_i\le k$ for each $i\le n$;
\item
$s_i\ne s_{i+1}$ for each $i < n$;
\item
$S$ contains no subsequence $xyxyxy\ldots$ of length $p+2$ for any $x\ne y$.
\end{itemize}
By a \emph{subsequence} of $S$ we mean any sequence $s_{i_1}s_{i_2}\ldots s_{i_m}$
with $i_1<i_2<\ldots <i_m$.
The maximum length of a $(k,p)$-Davenport-Schinzel sequence will be denoted
by $\lambda_p(k)$. We are interested in the particular case of $p=4$.

We inductively define a family of functions over positive integers:
$$
\begin{array}{llll}
A_1(n)&=&2n&n\ge1,\\
A_k(1)&=&2&k\ge1,\\
A_k(n)&=&A_{k-1}(A_k(n-1))&n\ge2,\,k\ge2.
\end{array}
$$
\emph{Ackermann's function} is defined by $A(n)=A_n(n)$.
This function grows faster than any primitive recursive function.
The inverse of Ackermann's function is defined by
$\alpha(n)=\min\setdef{t\ge1}{A(t)\ge n}$.

Agarwal, Sharir, and Shor \cite{ASS} proved that $\lambda_4(k)=O(k2^{\alpha(k)})$.
Note that $\alpha(n)$ grows very slowly, e.g., $\alpha(n)\le4$ for
all $n$ up to $A(4)$, which is the exponential tower of twos of height 65536. 
Thus, the bound for $\lambda_4(k)$ is nearly linear in~$k$.

Sometimes it will be convenient to identify a sequence $S=s_1\ldots s_n$ with all 
its cyclic shifts. This way $s_js_ns_1s_i$, where $i<j$,
is a subsequence of $S$. In such circumstances
we will call a sequence \emph{circular}.
Subsequences of $S$ will be
regarded also as circular sequences. Note that the set of all circular
subsequences is the same for $S$ and any of its shifts.
The length of $S$ will be denoted by~$|S|$.

\begin{lemma}\label{lem:circle}
Let $k,s\ge1$ and $S^{k,s}$ be the circular sequence consisting of $s$ successive
blocks of the form $12\ldots k$. 
\begin{bfenumerate}
\item
Suppose that $S$ is a subsequence of $S^{k,s}$ with no
4-subsubsequence of the form $xyxy$, where $x\ne y$. Then $|S|<k+s$.
\item
Suppose that $S$ is a subsequence of $S^{k,s}$ with no
6-subsubsequence of the form $xyxyxy$, where $x\ne y$. Then 
$|S|<\lambda_4(k)+s\le O(k2^{\alpha(k)})+s$.
\end{bfenumerate}
\end{lemma}

\begin{proof}
{\sl 1.}
We proceed by double induction on $k$ and $s$.
The base case where $k=1$ and $s$ is arbitrary is trivial.
Let $k\ge2$ and consider a subsequence $S$ with no forbidden subsubsequence. 
If each of the $k$ elements occurs in $S$ at most once, then
$|S|\le k$ and the claimed bound is true. Otherwise, without loss of generality 
we suppose that $S$ contains $\ell\ge2$ occurrences of $k$. 
Let $A_1,\ldots,A_\ell$ (resp.\ $B_1,\ldots,B_\ell$) 
denote the parts of $S$ (resp.\ $S^{k,s}$) between these $\ell$ elements.
Thus, $|S|=\ell+\sum_{i=1}^\ell|A_i|$. 

Denote the number of elements with at least one occurrence in $A_i$ by $k_i$.
Each element $x$ occurs in at most one of the $A_i$'s because otherwise
$S$ would contain a subsequence $xkxk$. It follows that $\sum_{i=1}^\ell k_i\le k-1$.
Note that, if we append $B_i$ with an element $k$,
it will consist of blocks $12\ldots k$. Denote the number of these blocks
by $s_i$ and notice the equality $\sum_{i=1}^\ell s_i=s$.
Since $A_i$ has no forbidden subsequence, we have $|A_i|\le k_i+s_i-1$.
If $k_i\ge1$, this follows from the induction assumption because
$A_i$ can be regarded a subsequence of $S^{k_i,s_i}$.
If $k_i=0$, this is also true because then $|A_i|=0$.
Summarizing, we obtain $|S|\le\ell+\sum_{i=1}^\ell(k_i+s_i-1)\le\ell+(k-1)+s-\ell<k+s$.

{\sl 2.}
Let $S'$ be obtained from $S$ by shrinking each block $z\ldots z$
of the same elements to $z$. Since $S'$ is a $(k,4)$-Davenport-Schinzel 
sequence, we have $|S'|\le\lambda_4(k)$. Note now that any two elements
neighboring in a shrunken block are at distance at least $k-1$ in $S^{k,s}$.
It easily follows that the total number of elements deleted in $S$ is less
than~$s$.
\end{proof}

\begin{theorem}\label{thm:fix}
Let $H_n$ be an arbitrary graph in $\calH_n$. For any $X$ in weakly convex position we have
$$\fixx{H_n}<3\sqrt n.$$ 
\end{theorem}

\begin{proof}
By part 1 of Lemma \ref{lem:fixxC}, it suffices to show that $C(X)<2k$
for any set $X$ of $n=k^2$ points on the boundary $\Gamma$ of a convex body.
Let $\colX$ be the interweaving $k$-coloring of $X$ where the colors appear
along $\Gamma$ in the circular sequence $S^{k,k}$ as in Lemma \ref{lem:circle}.
Suppose that $Y$ is a clustered subset of $X$. Note that there are no
two pairs $\{y_1,y_2\}\subset Y\cap X_i$ and $\{y'_1,y'_2\}\subset Y\cap X_j$, $i\ne j$,
with intersecting segments $[y_1,y_2]$ and $[y'_1,y'_2]$. This means that
the subsequence of $S^{k,k}$ induced by $Y$ does not contain any pattern
$ijij$. By part 1 of Lemma \ref{lem:circle}, we have $|Y|<2k$ and, hence,
$C(X,\colX)<2k$ as required.
\end{proof}

\begin{remark}\rm
With a little more care, we can improve the constant factor
in Theorem \ref{thm:fix} by proving that $\fixx{H_n}\le2\sqrt n+1$
for any $X$ in weakly convex position.
\end{remark}

The rest of this section is devoted to the star forest $kS_k$.
This sequence of graphs is of especial interest because this is the only
example of graphs for which we know that $\fix G=O(\sqrt n)$ but are currently able
to prove neither that $\FIX G=o(n)$ nor that $\fixx G=O(\sqrt n)$ for $X$
in weakly convex position.

The first part of the forthcoming Theorem~\ref{thm:stars} restates~\cite[Theorem 5]{Bose}
(see \refeq{stars} in Section \ref{ss:prior})
with a worse factor in front of $\sqrt n$; we include it for an expository
purpose. Somewhat surprisingly, the proof of this part is based on  part~1 of
Lemma~\ref{lem:circle}, which we already used to prove Theorem \ref{thm:fix}.
The second part, which is of our primary interest, requires a more delicate
analysis based on part 2 of Lemma~\ref{lem:circle}.

\begin{theorem}\label{thm:stars}
Let $kS_k$ denote the star forest with $n=k^2$ vertices.
For every integer $k\geq 2$ we have
\begin{bfenumerate}
\item
$\fixx{kS_k}<7\sqrt n$ for any collinear $X$;
\item
$\fixx{kS_k}=O(\sqrt n 2^{\alpha(\sqrt n)})$ for any $X$ in weakly convex position.
\end{bfenumerate}
\end{theorem}

\begin{proof}
Denote $V=V(kS_k)$.
Let $V=\bigcup_{i=1}^kV_i\cup C$, where each $V_i$ consists of all $k-1$ leaves
in the same star component and $C$ consists of all $k$ central vertices.

{\sl 1.}
Suppose that $X$ consists of points $x_1,\ldots,x_n$ lying on a line $\ell$ in this order.
Consider a drawing $\pi\function VX$ such that
\begin{equation}\label{eq:pi}
\begin{array}{rcl}
\pi(V_i)&=&\{x_i,x_{i+k},x_{i+2k},\ldots,x_{i+(k-2)k}\}\mathrm{\ for\ each\ }i\le k,\\
\pi(C)&=&\{x_{n-k+1},x_{n-k+2},\ldots,x_n\}.
\end{array}
\end{equation}
Let $\pi'$ be a crossing-free straight line redrawing of $kS_k$.
We have to estimate the number of fixed vertices, i.e., those vertices
participating in 
$F=\{ \hspace{0.5mm} \pi(v) : \hspace{0.5mm} v\in V,\ \pi(v)=\pi'(v)\}$.
For this purpose we split $F$ into four parts: $F=A\cup B\cup D\cup E$ where
$A$ (resp.\ $B$; $D$) consists of the fixed leaves adjacent to central vertices
located in $\pi'$ above $\ell$ (resp.\ below $\ell$; on $\ell$)
and $E$ consists of the fixed central vertices.

Trivially, $|E|\le k$ and it is easy to see that $|D|\le2k$. Let us estimate $|A|$ and $|B|$.
Label each $x_m$ by the index $i$ for which $x_m\in\pi(V_i)$ and view $x_1x_2\ldots x_{n-k}$
as the sequence $S^{k,k-1}$ defined in Lemma \ref{lem:circle}. 
Let $S$ be the subsequence induced by the points in $A$.
Note that $S$ does not contain any subsequence $ijij$ because
otherwise we would have an edge crossing in $\pi'$ (see Fig.~\ref{fig:cross}).
By part~1 of Lemma \ref{lem:circle}, we have $|A|=|S|<2k$. The same applies to $B$.
It follows that $|F|=|A|+|B|+|D|+|E|<7k$, as claimed.

\begin{figure}
\centerline{\includegraphics{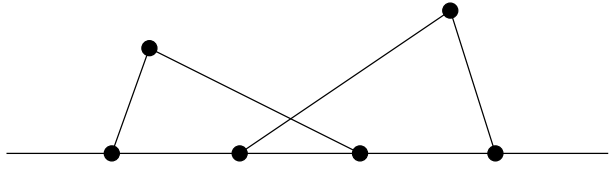}}
\caption{Proof of part 1 of Theorem \protect\ref{thm:stars}: an $ijij$-subsequence in $A$.}
\label{fig:cross}
\end{figure}

{\sl 2.}
Let $X$ be a set of $n=k^2$ points
on the boundary $\Gamma$ of a convex plane body $P$. 
It is known that the boundary of a convex plane body is a rectifiable curve and,
therefore, we can speak of the length of $\Gamma$ or its arcs.
Clearly, the convex body $P$ plays a nominal role and can be varied once $X$
is fixed.
Thus, to avoid unnecessary technical complications in the forthcoming
argument, without loss of generality we can suppose that the boundary
curve $\Gamma$ contains only a finite number of (maximal) straight line segments.
In particular, we can suppose that $\Gamma$ contains no straight line segment at all
if $X$ is in ``strictly'' convex position.

We will use the following terminology. A \emph{chord} is a straight line
segment whose endpoints lie on $\Gamma$. An \emph{arrow} is a directed chord
with one endpoint called \emph{head} and the other called \emph{tail}.
Call an arrow a \emph{median} if its endpoints split $\Gamma$ into arcs
of equal length. Fix the ``clockwise'' order of motion along $\Gamma$ and
color each non-median arrow in one of two colors, red if the shortest way
along $\Gamma$ from the tail to the head is clockwise and blue if it is
counter-clockwise.

Given a point $a$ outside $P$, we define \emph{quiver} $Q_a$ as follows.
For each line going through $a$ and intersecting $\Gamma$ in exactly two points,
$h$ and $t$, the $Q_a$ contains the arrow $th$ directed so that the head is closer
to $a$ than the tail. 

Given a non-median arrow $th$, we will denote the shorter component of
$\Gamma\setminus\{t,h\}$ by $\Gamma[t,h]$.
Our argument will be based on the following elementary fact.

\begin{claim}\label{cl:}
Let arrows $th$ and $t'h'$ be in the same quiver $Q$ and have the same color.
Suppose that $\Gamma[t',h']$ is shorter than $\Gamma[t,h]$. Then both $t'$ and $h'$
lie in $\Gamma[t,h]$.
\end{claim}

\begin{figure}
\centerline{\includegraphics{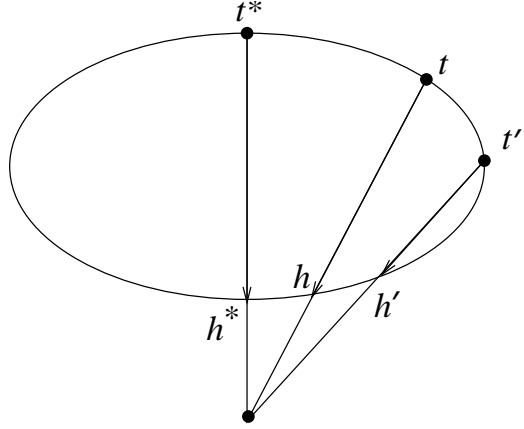}}
\caption{Proof of Claim \protect\ref{cl:}.}
\label{fig:quiver}
\end{figure}

\begin{subproof}
Let $t^*h^*$ be the median in $Q$. Since $th$ and $t'h'$ are of the same color,
the four points $t,h,t',h'$ are in the same component of $\Gamma\setminus\{t^*,h^*\}$.
The claim easily follows from the fact that the chords $th$ and $t'h'$ do not cross
(see Fig.~\ref{fig:quiver}).
\end{subproof}

After these preliminaries, we begin with the proof.
Let $x_1,\ldots,x_n$ be a listing of points in $X$ along $\Gamma$. 
Fix $\pi$ to be an arbitrary map satisfying \refeq{pi}.
Let $\pi'$ be a crossing-free redrawing of $kS_k$.
Look at the edges in $\pi'$ with one endpoint $\pi'(v)$ on $\Gamma$ 
and the other endpoint elsewhere. 
Perturbing $\pi'$ a little at the positions not lying on $\Gamma$ (and using the regularity
assumption made about $\Gamma$), we can ensure that
\begin{enumerate}
\item
any such edge intersects $\Gamma$ 
in at most two points, including $\pi'(v)$
(this is automatically true if $\Gamma$ contains no straight line segment);
\item
if an edge intersects $\Gamma$ in two points, 
it splits $\Gamma$ into components having different lengths.
\end{enumerate}

Assume that $\pi'$ meets these conditions.
Let $v$ be a leaf adjacent to a central
vertex $c$. Suppose that $\pi'(v)\in\Gamma$, $\pi'(c)\notin P$, and the segment
$\pi'(v)\pi'(c)$ crosses $\Gamma$ at a point $h\ne\pi'(v)$. By Condition 2,
the arrow $\pi'(v)h$ is not a median and hence colored in red or blue. 
We color each such $\pi'(v)$ in red or blue correspondingly.

Now we split the set of fixed vertices $F$ into five parts. 
Let $E$ consist of the fixed central vertices,
$I$ (resp.\ $O$) consist of those fixed leaves such that the edges emanating
from them are completely inside (resp.\ outside) $P$, and
$R$ (resp.\ $B$) consist of the red (resp.\ blue) fixed leaves.
By Condition 1, we have $F=E\cup I\cup O\cup R\cup B$.

Trivially, $|E|\le k$. Similarly to the proof of the first part of the theorem,
notice that the subsequences of $S^{k,k-1}$ corresponding to $I$ and $O$
do not contain $ijij$-subsubsequences. By part~1 of Lemma \ref{lem:circle},
we have $|I|<2k$ and $|O|<2k$. 

Finally, consider the subsequence $S$ of $S^{k,k-1}$ corresponding to $R$
and show that it does not contain any $ijijij$-subsubsequence.
Assume, to the contrary, that such a subsubsequence exists.
This means that $x_1\ldots x_{n-k}$ contains two interchanging subsequences
$a_1a_2a_3$ and $b_1b_2b_3$ whose elements belong to two different star components
of $\pi'$, with central vertices $a$ and $b$, respectively. 
Since $a_1,a_2,a_3$ are red, Claim \ref{cl:} implies that, say, $a_2$ and $a_3$
lie on the shorter arc of $\Gamma$ cut off by the edge $aa_1$ (see Fig.~\ref{fig:ababab}).

\begin{figure}
\centerline{\includegraphics{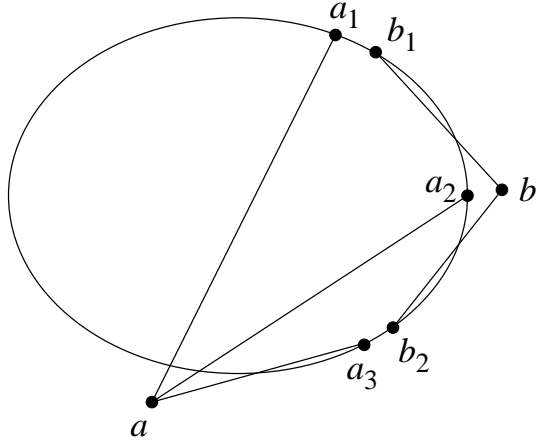}}
\caption{Proof of part 2 of Theorem \protect\ref{thm:stars}: 
impossibility of an $ijijij$-subsequence in $R$.}
\label{fig:ababab}
\end{figure}

Without loss of generality, let $b_1$ be between $a_1$ and $a_2$ and $b_2$
be between $a_2$ and $a_3$. Since $b_1$ and $b_2$ are red and $\pi'$ is
crossing-free, it must be the case that $bb_1$ intersects $\Gamma[a_1,a_2]$
and $bb_2$ intersects $\Gamma[a_2,a_3]$ (in another point). This makes a
contradiction with Claim \ref{cl:}.

Thus, $S$ is $ijijij$-free and, by part~2 of Lemma \ref{lem:circle}, we have
$|R|=|S|\le O(k2^{\alpha(k)})$. All the same applies to $B$.
Summarizing, we see that $|F|=|E|+|I|+|O|+|R|+|B|\le O(k2^{\alpha(k)})$, as claimed.
\end{proof}

\break

\section{Open problems}\label{s:open}
\mbox

\que
Can the parameters $\fix G$ and $\FIX G$ be far apart from each other
for some planar graphs?
Say, is it possible that for infinitely many graphs we have $\FIX G\ge n^\epsilon\fix G$
with a constant $\epsilon>0$?

\que
Lemma \ref{lem:CX} states an upper bound $C(X)=O(n/\log n)$ for any
set $X$ of $n=k^2$ points in the plane. A trivial lower bound is $C(X)\ge\sqrt n$.
How to make the gap closer? By Lemma \ref{lem:fixxC}, this way we could show either
that $\FIX{H_n}$ is close to $\fix{H_n}$ or that $\FIX{kS_k}$ is far from $\fix{kS_k}$.

\que
Find upper bounds on $\FIX G$, at least $\FIX G=o(n)$, for the cycle $C_n$,
the star forest $kS_k$, and the uniform binary tree.
Recall that upper bounds on $\fix G$ for these graphs are obtained
in \cite{PTa,Bose,Cib}, respectively 
(the uniform binary tree is just a particular instance
of the class of graphs with logarithmic vertex degrees and diameter treated in~\cite{Cib}).

\que
Let $\Fix G$ denote the maximum of $\fixx G$ over $X$ in weakly convex position.
Obviously, $\fix G\le\Fix G\le\FIX G$. Note that the first inequality can be strict: 
for example, $\fix{K_4}=2$ while $\Fix{K_4}=3$ for the tetrahedral graph.
Is it true that $\Fix G=O(\fix G)$? Currently we cannot prove this even for
graphs $G=kS_k$, cf.\ Theorem~\ref{thm:stars}.

\que
By Theorem \ref{thm:fixlfix}, for every $G$ we have $\fix G=\fixx G$ for some collinear
$X$. Does this equality hold for \emph{every} collinear $X$?
This question is related to the discussion in \cite[Section 5.1]{RVe}.

\subsection*{Acknowledgements}
We thank anonymous referees for their very careful reading of the manuscript
and suggesting several corrections and amendments.

\end{document}